\documentclass[aip,apl,twocolumn,groupedaddress,10pt]{revtex4-1}
\usepackage{amsmath}
\usepackage{graphicx}
\usepackage{color}

\begin{document}

\title{Theory of thermionic emission from a two-dimensional conductor and its application to a graphene-semiconductor Schottky junction}
\author{Maxim Trushin}
\affiliation{Centre for Advanced 2D Materials, National University of Singapore, 6 Science Drive 2, Singapore 117546}

\date{\today}

\begin{abstract}
The standard theory of thermionic emission developed 
for three-dimensional semiconductors does not apply to
two-dimensional materials even for making qualitative predictions
because of the vanishing out-of-plane quasiparticle velocity.
This study reveals the fundamental origin of the out-of-plane charge carrier motion 
in a two-dimensional conductor due to the finite quasiparticle lifetime and huge uncertainty of
the out-of-plane momentum. The theory is applied to  
a Schottky junction between graphene and a bulk semiconductor
to derive a thermionic constant, which, in contrast to the 
conventional Richardson constant, is determined by the 
Schottky barrier height and Fermi level in graphene.

\end{abstract}

\maketitle

\section{Introduction}

\begin{figure}
 \includegraphics[width=\columnwidth]{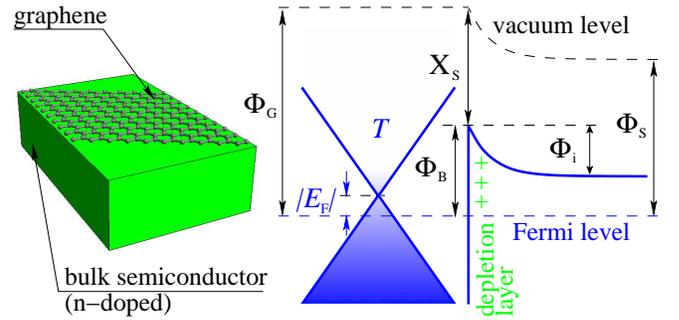}
 \caption{Graphene placed on top of a bulk n-doped semiconductor forms a Schottky junction.
Here, $\Phi_G$ and $\Phi_S$ are the work functions of graphene and the semiconductor, respectively, $\Phi_G > \Phi_S$ for Schottky contacts,
$E_F$ is the Fermi energy level counted from the band crossing point in graphene ($E_F <0$ in the configuration shown),
$T$ is the carrier temperature in graphene,
$X_S$ is the semiconductor affinity (independent of $E_F$),
$\Phi_i$ is the built-in potential created by the immobile ionized donors (green crosses), $\Phi_B$ is the Schottky barrier.
The in-plane band structure is shown for carriers in graphene.
For the bulk semiconductor only the bottom of the conduction band is shown.
The gradient filling shows the relative population of the bands due to finite temperature.
}
 \label{fig1}
\end{figure}

Microelectronic devices employing thermionic emission over the Schottky barrier between graphene and silicon
\cite{PRX2012tongay,liang2016modified,yang2012graphene,sinha2014ideal,chen2011graphene,song2015role,goykhman2016chip,liu2014quantum,li2016high,zhu2015tio,an2013tunable,an2013metal,an2015forward,jiao2016high,lin2015correlation,yim2013characterization,parui2014temperature,wang2013high,amirmazlaghani2013graphene,lv2013high,chen2015high,riazimehr2016spectral,riazimehr2017high,shen2017high,di2017hybrid,C7NR09591K}
or another bulk semiconductor
\cite{PRX2012tongay,liang2016modified,kim2016tunable,tomer2015carrier,kumar2016enhanced,mills2015direct,khurelbaatar2015temperature,poudel2017cross}
have experienced a boom over the last few years, see Refs. \onlinecite{di2016graphene,xu2016interface,xu2016contacts} for the most recent reviews.
Graphene-semiconductor Schottky junctions possess rectification properties 
\cite{PRX2012tongay,sinha2014ideal,yang2012graphene,chen2011graphene,an2015forward,lin2015correlation,kim2016tunable,tomer2015carrier,kumar2016enhanced,khurelbaatar2015temperature}
and can be used in photodetection
\cite{goykhman2016chip,liu2014quantum,li2016high,zhu2015tio,an2013tunable,an2013metal,an2015forward,di2016tunable,wang2013high,amirmazlaghani2013graphene,lv2013high,chen2015high,riazimehr2016spectral,riazimehr2017high,shen2017high,di2017hybrid,C7NR09591K}
as well as in solar energy harvesting.\cite{song2015role,jiao2016high,yawei2015two,olawole2018theoretical,miao2012high,an2013optimizing,li2015carbon}
The Schottky barrier height, $\Phi_B$,  is determined by the difference between the work function of graphene and semiconductor affinity,
see Fig. \ref{fig1}. Since the work function depends on the Fermi energy $E_F$ (which is tunable in graphene by an external electric field),
the barrier height depends on the bias voltage across the junction.\cite{PRX2012tongay,di2016graphene}
The thermionic current density through an ideal Schottky junction (i.e. no thermionic field emission, no series resistance etc.) then reads \cite{di2016graphene,crowell1966current}
\begin{equation}
 \label{I(V)}
 j=j_0 \left({\mathrm e}^{\frac{eV}{k_B T}} - 1 \right),
\end{equation}
where $e$ is the elementary charge, $k_B$ is the Boltzmann constant, $V$ is the bias voltage, and $T$ is the electron temperature.
The reverse saturation current density, $j_0$, flows across the junction
when a reverse bias voltage ($V<0$) pushes the electrons from graphene over the Schottky barrier to the semiconductor side.
The goal of this Letter is to derive $j_0$ for a two-dimensional (2D) conductor and apply the obtained formula to the 
graphene-semiconductor junction shown in Fig. \ref{fig1}.

The theoretical difficulties in modeling thermionic emission from 2D materials have been recognized only recently.
\cite{sinha2014ideal,liang2016modified,varonides2016combined,liang2015electron,ang2016current,ang2017generalized,upadhyay2017photo,misra2017thermionic,wei2013electron}
Indeed, thermionic emission from a surface of a bulk material is due to
(i) electron energy high enough to overcome the work function difference at the interface,
and (ii) non-zero electron velocity normal to the surface.
Electron kinetic energy can be controlled equally well in three-dimensional (3D)
as well as in  2D conductors by heating.
However, electrons in a 2D conductor (like graphene, which consists of only a surface) do not possess
an out-of-plane velocity and it is not clear where it should come from.
In attempt to circumvent this issue, Sinha and Lee have introduced 
an empirical carrier injection rate \cite{sinha2014ideal} (see also Refs. \onlinecite{wei2013electron,varonides2016combined}) whereas
Liang and Ang \cite{liang2015electron} have assumed a certain energy dispersion for the out-of-plane carrier motion,
see also Refs. \onlinecite{liang2016modified,ang2016current,ang2016current,upadhyay2017photo,misra2017thermionic,ang2017generalized} for further elaboration of their approach.
Despite their reasonable agreement with experiments,\cite{sinha2014ideal,liang2016modified} the models introduced until now
are not self-contained (i.e. they require an external parameter) and,
most importantly, do not explain {\em why} the 2D carriers do actually move out of plane.
Hence, the fundamental origin of the out-of-plane carrier velocity in thermionic emission from a 2D conductor is still unclear.
Here, I fill this gap using the concept of a hot electron liquid confined in a 2D plane.

\section{Quasiparticle lifetime and out-of-plane velocity}
The quasiparticle concept is a cornerstone of the electron liquid theory at finite temperatures:
Any given charge carrier does not stay in its state forever but may fall down or be excited to any other empty state
provided by energy and momentum conservation.
Hence, a quasiparticle state with energy near the Fermi level $E_F$
possess a finite lifetime given by\cite{abrikosov1988fundamentals} $\tau_{E_F} \sim \hbar E_F/(k_B T)^2$.
This estimation is valid for any normal metal, no matter 2D or 3D,\cite{lucas2018hydrodynamics}
as long as  $E_F\gg k_B T$. In intrinsic graphene, however,
the Fermi surface shrinks to a single point that results
in a 2D Dirac liquid \cite{lucas2018hydrodynamics} with the quasiparticle lifetime $\tau_0 \sim \hbar/(k_B T)$.
This lifetime can be seen as a mean time between the quasiparticle 
creation and annihilation events thanks to a time-independent perturbation.
Within this lifetime, $\tau$, a quasiparticle may acquire a certain perturbation-independent energy difference $\Delta E$
suggested by the energy uncertainty relation $\Delta E = \hbar/\tau$.
In contrast to $\Delta E$, the out-of-plane momentum uncertainty $\Delta p_z \sim \hbar /\Delta z$ is a constant determined solely
by the single atomic layer thickness $\Delta z$ being of the order of $1\,\mathring{\mathrm{A}}$.
Thanks to the finite $\Delta p_z$, a quasiparticle may acquire 
a finite out-of-plane velocity, $v_z = \Delta E/\Delta p_z$.\cite{landau2013quantum}
Assuming that the initial out-of-plane velocity is zero one can write the following
formula for the velocity $v_z$ that a quasiparticle acquires within the quasiparticle lifetime $\tau$:
\begin{equation}
\label{main}
v_z \Delta p_z \sim \hbar/\tau.
\end{equation}
Equation (\ref{main}) shows that the finite quasiparticle momentum uncertainty and lifetime are {\em both} necessary to correctly evaluate the out-of-plane velocity.
As long as the quasiparticle momentum may fluctuate within the uncertainty interval $\Delta p_z$, the quasiparticle acquires the finite velocity $v_z$.
The shorter quasiparticle lifetime is the higher velocity the quasiparticle has.
The out-of-plane velocity vanishes only at absolute zero temperature when $\tau\to\infty$.
In the case of zero momentum uncertainty (bulk limit, $\Delta z\to\infty$),
the out-of-plane velocity becomes disentangled from the energy uncertainty\cite{landau2013quantum} and is given by 
the standard relation $v_z=p_z/m^*$ in terms of an effective mass $m^*$.

The major merit of Eq. (\ref{main}) is to evaluate the out-of-plane quasiparticle velocity 
without the ill defined quasiparticle injection time or out-of-plane  effective mass. 
For quasiparticles in intrinsic graphene with $\tau=\tau_0$ and $T$ of a few hundreds of kelvins, one can estimate $v_z\sim 10^6$ cm/s,
i.e. it is two orders of magnitude lower than the in-plane Fermi velocity $v_F\sim 10^8$ cm/s.
If graphene is doped (say, $E_F\sim 0.2$ eV), then $\tau=\tau_{E_F}$ and the out-of-plane velocity is even lower, $v_z\sim 10^5$ cm/s.

\section{Thermionic emission}
Let us apply Eq. (\ref{main}) to an atomically thin conductor that forms a Schottky junction with a bulk semiconductor. 
The reverse saturation current density in Eq. (\ref{I(V)}) can be calculated by integrating the out-of-plane velocity
over the quasiparticle states with energies above the barrier as
\begin{equation}
\label{current1}
j_0 \sim eg_{sv}  \frac{v_z \Delta p_z }{2\pi\hbar} \int \frac{dp_y}{2\pi\hbar} 
\int \frac{dp_x}{2\pi\hbar} f^{(0)}_{E_{p_xp_y}},
\end{equation}
where $g_{sv}$ is the spin/valley degeneracy, $f^{(0)}_{E_{p_xp_y}}$ is the hot Fermi-Dirac distribution, $p_{x,y}$ ($p_z$) are
the in-plane (out-of-plane) components of the quasiparticle momentum, and, in contrast to the conventional approach,\cite{crowell1965richardson}
the integral over $p_z$ has been substituted by its uncertainty $\Delta p_z$.
Equation (\ref{current1}) is approximate and becomes applicable once the characteristic electron momentum 
(e.g. the Fermi momentum in metals) gets comparable with its uncertainty, or, in other words,
the quasiparticle de Broglie wavelength becomes comparable with the conductor thickness.
The integral over $p_z$ should be retained otherwise.


For a graphene-semiconductor Schottky junction shown in Fig. \ref{fig1}, we have $f^{(0)}_{E_{p_xp_y}}\sim \exp\left(\frac{E_F-E_{p_xp_y}}{k_BT}\right)$
at $k_B T \ll \Phi_B$ with $E_{p_xp_y}=v_F \sqrt{p_x^2 + p_y^2}$ for electrons above the barrier, $g_{sv}=4$, and Eq. (\ref{current1}) then reads
\begin{equation}
\label{current2}
j_0 \sim \frac{e}{\tau}\int\limits_{\Phi_B + E_F}^\infty 
\frac{dE E}{\pi^2\hbar^2 v_F^2}{\mathrm e}^{\frac{E_F-E}{k_BT}}, \quad k_B T \ll \Phi_B,
\end{equation}
where $\Phi_B=\Phi_{B0}-E_F$ with $\Phi_{B0}$ being the Schottky barrier height at $E_F=0$, see Fig. \ref{fig1}.
In the case of intrinsic graphene ($|E_F|\ll k_B T$, $\tau=\tau_0$) we have 
\begin{equation}
\label{current3}
j_0 \sim A^*_{G0} T^2{\mathrm e}^{-\frac{\Phi_B}{k_BT}}, \quad |E_F|\ll k_B T.
\end{equation}
Here, $A^*_{G0}=e(\Phi_{B0}+ k_B T)k_B^2/(\pi^2 \hbar^3 v_F^2)$ is the thermionic constant for intrinsic graphene.
Since  $k_B T$ is much lower than $\Phi_{B0}$, the former can be neglected in $A^*_{G0}$, and
the current density obeys the Richardson-Dushman law,\cite{dushman1930thermionic} where the reverse saturation current is 
given by $j_0=A^* T^2 \exp[-\Phi_B/(k_BT)]$ with $A^*=em^*k_B^2/(2\pi^2\hbar^3)$ being the Richardson constant.\cite{crowell1965richardson,dushman1930thermionic}
In contrast to the conventional Richardson constant,  $A^*_{G0}$ strongly depends on the barrier height $\Phi_{B0}$.
For $\Phi_{B0}$ of the order of $0.1$ eV (see Table 1 in Ref. \onlinecite{di2016graphene}), we estimate $A^*_{G0}$ to be about
$10\,\mathrm{A/cm^2/K^2}$ that is comparable with the Richardson constant $A^* = 120\, (m^*/m_0)\,\mathrm{A/cm^2/K^2}$,
where the effective-to-free electron mass ratio $m^*/m_0$ is of the order of $0.1$ for typical bulk metal-semiconductor junctions.\cite{missous1991richardson}
One could define an out-of-plane effective mass for carriers in graphene as $m^*v_F^2/2=\Phi_B$  
and utilize the standard formula for $A^*$. The Schottky barrier height plays therefore a role of the effective
electron mass in the conventional theory of thermionic emission. This makes sense because both 
the Schottky barrier and inertial mass resist the out-of-plane particle acceleration.

In the case of doped graphene ($|E_F|\gg k_B T$, $\tau=\tau_{E_F}$), Eq. (\ref{current2}) reads
\begin{equation}
\label{current4}
j_0 \sim  \frac{e(\Phi_{B0}+k_B T)  k_B^3 T^3}{\pi^2 \hbar^3 v_F^2 |E_F|}
{\mathrm e}^{-\frac{\Phi_B}{k_BT}}, \quad k_B T \ll |E_F|,
\end{equation}
where $E_F$ can be positive or negative depending on graphene doping, and
$T^3$-dependence indicates that the Richardson-Dushman law becomes invalid.
Nevertheless, one can formally define the temperature dependent thermionic constant as $A^*_G = A^*_{G0} \frac{k_B T}{|E_F|}$.
Hence, the thermionic constant is further reduced as compared with the intrinsic value $A^*_{G0}$
and could therefore partly explain why its actual value measured in graphene-silicon junctions is smaller than expected.\cite{sinha2014ideal}
The temperature dependence of $j_0$ given by Eq. (\ref{current4}) at $\Phi_{B0} \gg k_B T$ is similar to what has been predicted by Liang and Ang\cite{liang2015electron}
but, in contrast to their model,  $A^*_G$ is governed by $\Phi_{B0}$ and $E_F$, not just by the fundamental constants and Fermi velocity in graphene.





\section{Discussion}

\begin{figure}
 \includegraphics[width=0.92\columnwidth]{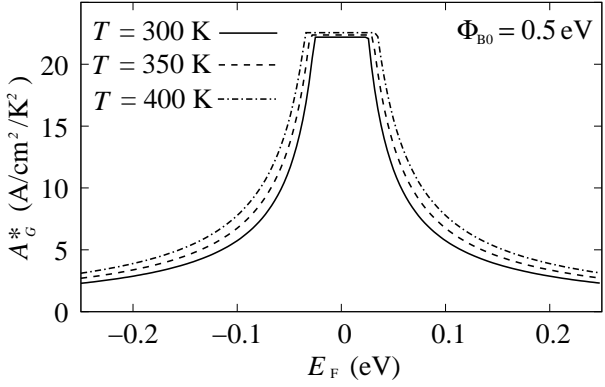}
 \caption{Schematic plot of the thermionic constant $A^*_G$ vs. the Fermi energy $E_F$, as defined below Eq. (\ref{current4}).
 Since $A^*_G$ diverges at $E_F=0$, it is substituted by the intrinsic value $A^*_{G0}$
 once $A^*_G$ becomes larger than $A^*_{G0}$. The intrinsic value is defined below Eq. (\ref{current3}).
 The Schottky barrier height at $E_F=0$ is $\Phi_{B0} \sim 0.5$ eV typical for graphene-semiconductor junctions.\cite{di2016graphene}
 The Fermi level depends on the bias voltage \cite{di2016graphene} that makes the thermionic constant bias-dependent as well. }
 \label{fig2}
\end{figure}

\begin{figure}
 \includegraphics[width=\columnwidth]{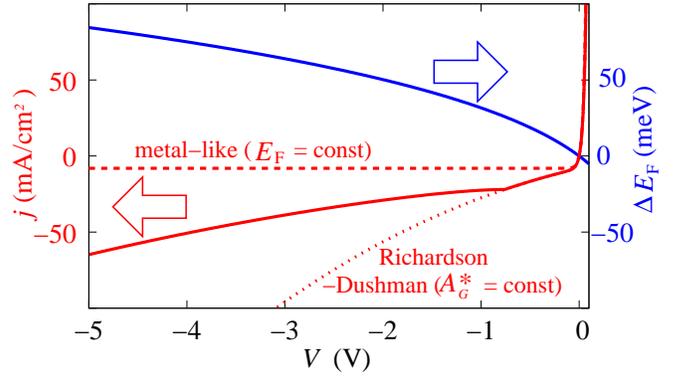}
 \caption{Typical current-voltage characteristics (solid red line)  plotted 
 for an intrinsic graphene/n-type semiconductor Schottky junction ($\Phi_{B0} \sim 0.5$ eV, $T\sim 300$ K) by means of Eq. (\ref{I(V)})
 with $j_0$ given by either Eqs. (\ref{current3}) or (\ref{current4}), depending on $E_F$.
 The dotted line corresponds to the conventional Richardson-Dushman model with the thermionic constant $A^*_G=A^*_{G0}$
 assumed to be independent of the Fermi energy $E_F$.
 The dashed line describes the metal-like model when the Fermi energy does not depend on the bias voltage $V$
 so that the saturation current, $j_0$, remains constant.
 The blue solid line is the (quasi-)Fermi energy shift in graphene due to the carrier density induced by the bias voltage.
 The parameters are discussed in the main text.
 }
 \label{fig3}
\end{figure}

The thermionic constant turns out to be strongly dependent on the Fermi energy
and offers an interesting opportunity to test the model predictions in real devices.
The constant increases while the Fermi level approaches the band crossing point in graphene
until it reaches the intrinsic value $A^*_{G0}$ at $|E_F|\ll k_B T$.
The trend is schematically shown in Fig. \ref{fig2} at $\Phi_{B0}\sim 0.5$ eV.
The saturation current, however, does not follow this trend since
the inverse proportion to $E_F$ is by far compensated for by the exponential dependence on $E_F$ via $\Phi_B$.
Anyway, the thermionic constant can easily be identified by plotting $j_0(E_F)$ on a logarithmic scale
and subtracting the trivial $\Phi_B/T$ term.\cite{missous1991richardson}

In contrast to conventional metal-semiconductor diodes, graphene-semiconductor junctions demonstrate 
a bias-driven increase of the reverse saturation current.\cite{PRX2012tongay,di2016graphene}
To be specific, let us consider the current-voltage characteristics of a junction between graphene and an n-doped semiconductor
assuming that the Fermi level intercepts the band crossing point in graphene at zero bias (see Fig. 17 in Ref. [\onlinecite{di2016graphene}]).
A bias voltage redistributes the positive and negative charge across the junction, resulting in two different (quasi)-Fermi levels
for carriers on the graphene and semiconductor sides.
The forward bias ($V>0$) lowers the positive charge of the depletion layer in the semiconductor,
hence reducing the built-in potential $\Phi_i$ at the interface as well as the Fermi energy in graphene.
In contrast, the reverse bias ($V<0$) extends the depletion layer to the semiconductor side increasing the built-in potential.
At the same time, the Fermi level increases in graphene because of the higher charge density induced there 
to mirror the excessive immobile donor charge of the depletion layer. The induced charge density can be 
calculated using the Schottky-Mott relationship as\cite{di2016graphene}
\begin{eqnarray}
 \nonumber
-e \Delta n_\mathrm{ind} &=& \sqrt{2\epsilon_S N_D \left(\Phi_i - eV - k_B T\right)} \\
&& - \sqrt{2\epsilon_S N_D \left(\Phi_i - k_B T\right)},
\label{dnind}
\end{eqnarray}
where $\epsilon_S\sim 10$ is the relative permittivity of a typical semiconductor, $N_D \sim 10^{16}$ cm$^{-3}$ is the 
donor concentration in the depletion layer, and $\Phi_i\sim 0.6$ eV is the built-in potential.
The typical parameter values are taken from Ref. [\onlinecite{PRX2012tongay}] where various graphene/n-type semiconductor junctions have been studied.
The Fermi energy difference due to the reverse bias can be written for graphene as $\Delta E_F = \hbar v_F \sqrt{-\pi \Delta n_\mathrm{ind}}$,
where the four-fold degeneracy has been taken into account in the relation between the Fermi wave vector and the carrier concentration.
The blue line in Fig. \ref{fig3} demonstrates that the reverse bias results in a transfer of negative charge 
to graphene large enough to increase $\Delta E_F$ by tens of meV at the voltages of a few volts, in accordance with the measurements of Ref. \onlinecite{PRX2012tongay}.
The upward shift of $E_F$ in graphene causes a reduction of the barrier height $\Phi_B$ 
facilitating electron transport across the junction and preventing the reverse current from saturating.
If graphene behaved like a regular metal with a very high density of states,
then the Fermi energy would not change much under the bias voltage resulting in a nearly constant $j_0$
and perfect saturation of the reverse current (see the red dashed line in Fig. \ref{fig3}).
Retaining the exponential dependence $j_0$ on $E_F$ but neglecting the substantial decrease
in the thermionic constant at higher $|E_F|$  
allows us to qualitatively reproduce the lack of the reverse current saturation observed in real graphene-semiconductor junctions\cite{PRX2012tongay}
(see the red dotted line in Fig. \ref{fig3}).
In fact, the inverse proportion of $j_0$ to $E_F$ read out from Eq. (\ref{current4}) partly compensates for this exponential trend.
The result is that the true reverse current values shown by the red solid line in Fig. \ref{fig3} lie between
the curves given by Eq. (\ref{current3}) with $E_F=0$ assumed to be independent of $V$ (the metal-like model, dashed curve) and
by the same equation with $E_F=\Delta E_F(V)$  (the Richardson-Dushman law, dotted curve).

The experimental verification of the model proposed is obviously 
of the utmost importance for graphene-based device applications as the reverse saturation current 
is one of the most essential parameters to characterize a Schottky diode.
The previous transport measurements\cite{PRX2012tongay} have employed samples fabricated from graphene grown on copper by chemical vapor deposition
and subsequently transfered onto a semiconductor. Graphene has been found to be p-doped with the Fermi energy $|E_F| \gtrsim 200$ meV being much higher
than $\Delta E_F$ provided by the bias voltage (see Fig. \ref{fig3}).
It was therefore impossible to observe the qualitative change of the saturation current behavior due to
the crossover between intrinsic ($|E_F| \ll k_B T$) and highly doped ($k_B T \ll |E_F|$) regimes.
Moreover, the reverse current seen in Fig. \ref{fig3} is much higher than the measured one\cite{PRX2012tongay}
because the Schottky barrier is almost $0.3$ eV lower in our case of intrinsic graphene.
To see the effects predicted above, one needs much cleaner samples with $|E_F| \lesssim 25$ meV at zero bias.
Nevertheless, the $T^3$-dependence in the saturation current for doped graphene predicted by Eq. (\ref{current4})
can be verified by using existing experimental data\cite{liang2016modified} for graphene contacts with Si, MoS$_2$, GaAs, and GaN.
Indeed, the measured current density plotted as $\ln(j_0/T^3)$ vs. $1/T$ 
demonstrates a linear dependence for any of the four bulk semiconductors considered.
This is a clear indication of consistency between the theory and experimental data available at the moment. 

Besides possible applications, the experiments with nearly intrinsic graphene 
might shed some light on the fundamental properties of a Dirac liquid
confined in a 2D plane (see Ref. \onlinecite{lucas2018hydrodynamics} for review).
Indeed, the most important ingredient of the model introduced above is the quasiparticle lifetime which is $\tau_0 \sim \hbar/(k_B T)$ in the intrinsic limit. 
Interestingly, an emission rate equal to $k_B T/\hbar$ has been obtained in Ref. \onlinecite{wei2013electron}
by fitting the thermionic constant for a 2D electron gas with the Richardson-Dushman law.
Here, we associate the emission rate with the inverted quasiparticle lifetime, which makes a lot of sense
as $\tau_0$ can be seen as the average time needed to repopulate a given state above the barrier after emission.
Once the Fermi energy is shifted away from the band crossing point in graphene,
the quasiparticle lifetime decreases rapidly and, as a consequence,
the thermionic constant is reduced.  Since the Fermi energy depends on the bias voltage, 
this effect should be seen in the current-voltage measurements.
Alternatively, the Fermi energy of graphene can be varied through appropriate gating.\cite{yang2012graphene,chen2012gate}

\section{Summary and Outlook}

The intimate relation between the quasiparticle emission rate and lifetime
is among the main findings of this work.
The model explains, above all, why the quasiparticles tightly confined
in a 2D plane still possess an out-of-plane velocity.
The thermionic emission mechanism in graphene-semiconductor junctions strongly differs from its bulk version.
The Richardson constant employed in the conventional model
is substituted by the thermionic constant $A^*_G$
that depends on the Fermi energy that is in turn tunable by an external electric field normal to the graphene plane. 

Graphene-semiconductor junctions should demonstrate efficient photocarrier injection with the excitation energy below the semiconductor bandgap.
The photocarriers excited in graphene are thermalized rapidly \cite{malic2012efficient,brida2013ultrafast} creating a high-energy distribution tail above the Schottky barrier,
hence providing excess energy necessary for thermionic emission across the interface.
In contrast to the conventional heterojunctions,\cite{wu1979carrier,kim2010momentum}
the out-of-plane momentum conservation is relaxed by the momentum uncertainty $\Delta p_z \sim \hbar /\Delta z$,
which, thanks to the small layer thickness $\Delta z$, is of the order of the first Brillouin zone size.
The interface disorder thereby does not much hamper the out-of-plane carrier transport between graphene and the semiconductor.
The injection rate is  governed  by the quasiparticle lifetime rather than the interface disorder.
Hence, the carriers do not behave like point-like particles along $z$-direction until they 
turn out to be on the semiconductor side with a well-defined momentum.
This peculiarity should be taken into account when designing optoelectronic devices
based on 2D materials.

\acknowledgments

I would like to thank Ilya Goykhman and Andrea Ferrari for turning my attention to the problem of thermionic emission from graphene.
Multiple discussions with Shi-Jun Liang are also acknowledged.
This work has been supported by the Director's Senior Research Fellowship from the Centre for Advanced 2D Materials at the National University of Singapore
(NRF Medium Sized Centre Programme R-723-000-001-281).

\bibliography{2Dto3Dthermionics.bib}

\begin{thebibliography}{60}%
\makeatletter
\providecommand \@ifxundefined [1]{%
 \@ifx{#1\undefined}
}%
\providecommand \@ifnum [1]{%
 \ifnum #1\expandafter \@firstoftwo
 \else \expandafter \@secondoftwo
 \fi
}%
\providecommand \@ifx [1]{%
 \ifx #1\expandafter \@firstoftwo
 \else \expandafter \@secondoftwo
 \fi
}%
\providecommand \natexlab [1]{#1}%
\providecommand \enquote  [1]{``#1''}%
\providecommand \bibnamefont  [1]{#1}%
\providecommand \bibfnamefont [1]{#1}%
\providecommand \citenamefont [1]{#1}%
\providecommand \href@noop [0]{\@secondoftwo}%
\providecommand \href [0]{\begingroup \@sanitize@url \@href}%
\providecommand \@href[1]{\@@startlink{#1}\@@href}%
\providecommand \@@href[1]{\endgroup#1\@@endlink}%
\providecommand \@sanitize@url [0]{\catcode `\\12\catcode `\$12\catcode
  `\&12\catcode `\#12\catcode `\^12\catcode `\_12\catcode `\%12\relax}%
\providecommand \@@startlink[1]{}%
\providecommand \@@endlink[0]{}%
\providecommand \url  [0]{\begingroup\@sanitize@url \@url }%
\providecommand \@url [1]{\endgroup\@href {#1}{\urlprefix }}%
\providecommand \urlprefix  [0]{URL }%
\providecommand \Eprint [0]{\href }%
\providecommand \doibase [0]{http://dx.doi.org/}%
\providecommand \selectlanguage [0]{\@gobble}%
\providecommand \bibinfo  [0]{\@secondoftwo}%
\providecommand \bibfield  [0]{\@secondoftwo}%
\providecommand \translation [1]{[#1]}%
\providecommand \BibitemOpen [0]{}%
\providecommand \bibitemStop [0]{}%
\providecommand \bibitemNoStop [0]{.\EOS\space}%
\providecommand \EOS [0]{\spacefactor3000\relax}%
\providecommand \BibitemShut  [1]{\csname bibitem#1\endcsname}%
\let\auto@bib@innerbib\@empty
\bibitem [{\citenamefont {Tongay}\ \emph {et~al.}(2012)\citenamefont {Tongay},
  \citenamefont {Lemaitre}, \citenamefont {Miao}, \citenamefont {Gila},
  \citenamefont {Appleton},\ and\ \citenamefont {Hebard}}]{PRX2012tongay}%
  \BibitemOpen
  \bibfield  {author} {\bibinfo {author} {\bibfnamefont {S.}~\bibnamefont
  {Tongay}}, \bibinfo {author} {\bibfnamefont {M.}~\bibnamefont {Lemaitre}},
  \bibinfo {author} {\bibfnamefont {X.}~\bibnamefont {Miao}}, \bibinfo {author}
  {\bibfnamefont {B.}~\bibnamefont {Gila}}, \bibinfo {author} {\bibfnamefont
  {B.~R.}\ \bibnamefont {Appleton}}, \ and\ \bibinfo {author} {\bibfnamefont
  {A.~F.}\ \bibnamefont {Hebard}},\ }\href {\doibase 10.1103/PhysRevX.2.011002}
  {\bibfield  {journal} {\bibinfo  {journal} {Phys. Rev. X}\ }\textbf {\bibinfo
  {volume} {2}},\ \bibinfo {pages} {011002} (\bibinfo {year}
  {2012})}\BibitemShut {NoStop}%
\bibitem [{\citenamefont {Liang}\ \emph {et~al.}(2016)\citenamefont {Liang},
  \citenamefont {Hu}, \citenamefont {Di~Bartolomeo}, \citenamefont {Adam},\
  and\ \citenamefont {Ang}}]{liang2016modified}%
  \BibitemOpen
  \bibfield  {author} {\bibinfo {author} {\bibfnamefont {S.-J.}\ \bibnamefont
  {Liang}}, \bibinfo {author} {\bibfnamefont {W.}~\bibnamefont {Hu}}, \bibinfo
  {author} {\bibfnamefont {A.}~\bibnamefont {Di~Bartolomeo}}, \bibinfo {author}
  {\bibfnamefont {S.}~\bibnamefont {Adam}}, \ and\ \bibinfo {author}
  {\bibfnamefont {L.~K.}\ \bibnamefont {Ang}},\ }in\ \href@noop {} {\emph
  {\bibinfo {booktitle} {Electron Devices Meeting (IEDM), 2016 IEEE
  International}}}\ (\bibinfo {organization} {IEEE},\ \bibinfo {year} {2016})\
  pp.\ \bibinfo {pages} {14--4}\BibitemShut {NoStop}%
\bibitem [{\citenamefont {Yang}\ \emph {et~al.}(2012)\citenamefont {Yang},
  \citenamefont {Heo}, \citenamefont {Park}, \citenamefont {Song},
  \citenamefont {Seo}, \citenamefont {Byun}, \citenamefont {Kim}, \citenamefont
  {Yoo}, \citenamefont {Chung},\ and\ \citenamefont {Kim}}]{yang2012graphene}%
  \BibitemOpen
  \bibfield  {author} {\bibinfo {author} {\bibfnamefont {H.}~\bibnamefont
  {Yang}}, \bibinfo {author} {\bibfnamefont {J.}~\bibnamefont {Heo}}, \bibinfo
  {author} {\bibfnamefont {S.}~\bibnamefont {Park}}, \bibinfo {author}
  {\bibfnamefont {H.~J.}\ \bibnamefont {Song}}, \bibinfo {author}
  {\bibfnamefont {D.~H.}\ \bibnamefont {Seo}}, \bibinfo {author} {\bibfnamefont
  {K.-E.}\ \bibnamefont {Byun}}, \bibinfo {author} {\bibfnamefont
  {P.}~\bibnamefont {Kim}}, \bibinfo {author} {\bibfnamefont {I.}~\bibnamefont
  {Yoo}}, \bibinfo {author} {\bibfnamefont {H.-J.}\ \bibnamefont {Chung}}, \
  and\ \bibinfo {author} {\bibfnamefont {K.}~\bibnamefont {Kim}},\ }\href@noop
  {} {\bibfield  {journal} {\bibinfo  {journal} {Science}\ }\textbf {\bibinfo
  {volume} {336}},\ \bibinfo {pages} {1140} (\bibinfo {year}
  {2012})}\BibitemShut {NoStop}%
\bibitem [{\citenamefont {Sinha}\ and\ \citenamefont
  {Lee}(2014)}]{sinha2014ideal}%
  \BibitemOpen
  \bibfield  {author} {\bibinfo {author} {\bibfnamefont {D.}~\bibnamefont
  {Sinha}}\ and\ \bibinfo {author} {\bibfnamefont {J.~U.}\ \bibnamefont
  {Lee}},\ }\href@noop {} {\bibfield  {journal} {\bibinfo  {journal} {Nano
  Letters}\ }\textbf {\bibinfo {volume} {14}},\ \bibinfo {pages} {4660}
  (\bibinfo {year} {2014})}\BibitemShut {NoStop}%
\bibitem [{\citenamefont {Chen}\ \emph {et~al.}(2011)\citenamefont {Chen},
  \citenamefont {Aykol}, \citenamefont {Chang}, \citenamefont {Levi},\ and\
  \citenamefont {Cronin}}]{chen2011graphene}%
  \BibitemOpen
  \bibfield  {author} {\bibinfo {author} {\bibfnamefont {C.-C.}\ \bibnamefont
  {Chen}}, \bibinfo {author} {\bibfnamefont {M.}~\bibnamefont {Aykol}},
  \bibinfo {author} {\bibfnamefont {C.-C.}\ \bibnamefont {Chang}}, \bibinfo
  {author} {\bibfnamefont {A.}~\bibnamefont {Levi}}, \ and\ \bibinfo {author}
  {\bibfnamefont {S.~B.}\ \bibnamefont {Cronin}},\ }\href@noop {} {\bibfield
  {journal} {\bibinfo  {journal} {Nano Letters}\ }\textbf {\bibinfo {volume}
  {11}},\ \bibinfo {pages} {1863} (\bibinfo {year} {2011})}\BibitemShut
  {NoStop}%
\bibitem [{\citenamefont {Song}\ \emph {et~al.}(2015)\citenamefont {Song},
  \citenamefont {Li}, \citenamefont {Mackin}, \citenamefont {Zhang},
  \citenamefont {Fang}, \citenamefont {Palacios}, \citenamefont {Zhu},\ and\
  \citenamefont {Kong}}]{song2015role}%
  \BibitemOpen
  \bibfield  {author} {\bibinfo {author} {\bibfnamefont {Y.}~\bibnamefont
  {Song}}, \bibinfo {author} {\bibfnamefont {X.}~\bibnamefont {Li}}, \bibinfo
  {author} {\bibfnamefont {C.}~\bibnamefont {Mackin}}, \bibinfo {author}
  {\bibfnamefont {X.}~\bibnamefont {Zhang}}, \bibinfo {author} {\bibfnamefont
  {W.}~\bibnamefont {Fang}}, \bibinfo {author} {\bibfnamefont {T.}~\bibnamefont
  {Palacios}}, \bibinfo {author} {\bibfnamefont {H.}~\bibnamefont {Zhu}}, \
  and\ \bibinfo {author} {\bibfnamefont {J.}~\bibnamefont {Kong}},\ }\href@noop
  {} {\bibfield  {journal} {\bibinfo  {journal} {Nano Letters}\ }\textbf
  {\bibinfo {volume} {15}},\ \bibinfo {pages} {2104} (\bibinfo {year}
  {2015})}\BibitemShut {NoStop}%
\bibitem [{\citenamefont {Goykhman}\ \emph {et~al.}(2016)\citenamefont
  {Goykhman}, \citenamefont {Sassi}, \citenamefont {Desiatov}, \citenamefont
  {Mazurski}, \citenamefont {Milana}, \citenamefont {de~Fazio}, \citenamefont
  {Eiden}, \citenamefont {Khurgin}, \citenamefont {Shappir}, \citenamefont
  {Levy} \emph {et~al.}}]{goykhman2016chip}%
  \BibitemOpen
  \bibfield  {author} {\bibinfo {author} {\bibfnamefont {I.}~\bibnamefont
  {Goykhman}}, \bibinfo {author} {\bibfnamefont {U.}~\bibnamefont {Sassi}},
  \bibinfo {author} {\bibfnamefont {B.}~\bibnamefont {Desiatov}}, \bibinfo
  {author} {\bibfnamefont {N.}~\bibnamefont {Mazurski}}, \bibinfo {author}
  {\bibfnamefont {S.}~\bibnamefont {Milana}}, \bibinfo {author} {\bibfnamefont
  {D.}~\bibnamefont {de~Fazio}}, \bibinfo {author} {\bibfnamefont
  {A.}~\bibnamefont {Eiden}}, \bibinfo {author} {\bibfnamefont
  {J.}~\bibnamefont {Khurgin}}, \bibinfo {author} {\bibfnamefont
  {J.}~\bibnamefont {Shappir}}, \bibinfo {author} {\bibfnamefont
  {U.}~\bibnamefont {Levy}},  \emph {et~al.},\ }\href@noop {} {\bibfield
  {journal} {\bibinfo  {journal} {Nano Letters}\ }\textbf {\bibinfo {volume}
  {16}},\ \bibinfo {pages} {3005} (\bibinfo {year} {2016})}\BibitemShut
  {NoStop}%
\bibitem [{\citenamefont {Liu}\ and\ \citenamefont
  {Kar}(2014)}]{liu2014quantum}%
  \BibitemOpen
  \bibfield  {author} {\bibinfo {author} {\bibfnamefont {F.}~\bibnamefont
  {Liu}}\ and\ \bibinfo {author} {\bibfnamefont {S.}~\bibnamefont {Kar}},\
  }\href@noop {} {\bibfield  {journal} {\bibinfo  {journal} {ACS Nano}\
  }\textbf {\bibinfo {volume} {8}},\ \bibinfo {pages} {10270} (\bibinfo {year}
  {2014})}\BibitemShut {NoStop}%
\bibitem [{\citenamefont {Li}\ \emph {et~al.}(2016)\citenamefont {Li},
  \citenamefont {Zhu}, \citenamefont {Du}, \citenamefont {Lv}, \citenamefont
  {Zhang}, \citenamefont {Li}, \citenamefont {Yang}, \citenamefont {Yang},
  \citenamefont {Li}, \citenamefont {Wang} \emph {et~al.}}]{li2016high}%
  \BibitemOpen
  \bibfield  {author} {\bibinfo {author} {\bibfnamefont {X.}~\bibnamefont
  {Li}}, \bibinfo {author} {\bibfnamefont {M.}~\bibnamefont {Zhu}}, \bibinfo
  {author} {\bibfnamefont {M.}~\bibnamefont {Du}}, \bibinfo {author}
  {\bibfnamefont {Z.}~\bibnamefont {Lv}}, \bibinfo {author} {\bibfnamefont
  {L.}~\bibnamefont {Zhang}}, \bibinfo {author} {\bibfnamefont
  {Y.}~\bibnamefont {Li}}, \bibinfo {author} {\bibfnamefont {Y.}~\bibnamefont
  {Yang}}, \bibinfo {author} {\bibfnamefont {T.}~\bibnamefont {Yang}}, \bibinfo
  {author} {\bibfnamefont {X.}~\bibnamefont {Li}}, \bibinfo {author}
  {\bibfnamefont {K.}~\bibnamefont {Wang}},  \emph {et~al.},\ }\href@noop {}
  {\bibfield  {journal} {\bibinfo  {journal} {Small}\ }\textbf {\bibinfo
  {volume} {12}},\ \bibinfo {pages} {595} (\bibinfo {year} {2016})}\BibitemShut
  {NoStop}%
\bibitem [{\citenamefont {Zhu}\ \emph {et~al.}(2015)\citenamefont {Zhu},
  \citenamefont {Zhang}, \citenamefont {Li}, \citenamefont {He}, \citenamefont
  {Li}, \citenamefont {Guo}, \citenamefont {Zang}, \citenamefont {Wang},
  \citenamefont {Xie}, \citenamefont {Li} \emph {et~al.}}]{zhu2015tio}%
  \BibitemOpen
  \bibfield  {author} {\bibinfo {author} {\bibfnamefont {M.}~\bibnamefont
  {Zhu}}, \bibinfo {author} {\bibfnamefont {L.}~\bibnamefont {Zhang}}, \bibinfo
  {author} {\bibfnamefont {X.}~\bibnamefont {Li}}, \bibinfo {author}
  {\bibfnamefont {Y.}~\bibnamefont {He}}, \bibinfo {author} {\bibfnamefont
  {X.}~\bibnamefont {Li}}, \bibinfo {author} {\bibfnamefont {F.}~\bibnamefont
  {Guo}}, \bibinfo {author} {\bibfnamefont {X.}~\bibnamefont {Zang}}, \bibinfo
  {author} {\bibfnamefont {K.}~\bibnamefont {Wang}}, \bibinfo {author}
  {\bibfnamefont {D.}~\bibnamefont {Xie}}, \bibinfo {author} {\bibfnamefont
  {X.}~\bibnamefont {Li}},  \emph {et~al.},\ }\href@noop {} {\bibfield
  {journal} {\bibinfo  {journal} {Journal of Materials Chemistry A}\ }\textbf
  {\bibinfo {volume} {3}},\ \bibinfo {pages} {8133} (\bibinfo {year}
  {2015})}\BibitemShut {NoStop}%
\bibitem [{\citenamefont {An}\ \emph {et~al.}(2013{\natexlab{a}})\citenamefont
  {An}, \citenamefont {Liu}, \citenamefont {Jung},\ and\ \citenamefont
  {Kar}}]{an2013tunable}%
  \BibitemOpen
  \bibfield  {author} {\bibinfo {author} {\bibfnamefont {X.}~\bibnamefont
  {An}}, \bibinfo {author} {\bibfnamefont {F.}~\bibnamefont {Liu}}, \bibinfo
  {author} {\bibfnamefont {Y.~J.}\ \bibnamefont {Jung}}, \ and\ \bibinfo
  {author} {\bibfnamefont {S.}~\bibnamefont {Kar}},\ }\href@noop {} {\bibfield
  {journal} {\bibinfo  {journal} {Nano Letters}\ }\textbf {\bibinfo {volume}
  {13}},\ \bibinfo {pages} {909} (\bibinfo {year}
  {2013}{\natexlab{a}})}\BibitemShut {NoStop}%
\bibitem [{\citenamefont {An}\ \emph {et~al.}(2013{\natexlab{b}})\citenamefont
  {An}, \citenamefont {Behnam}, \citenamefont {Pop},\ and\ \citenamefont
  {Ural}}]{an2013metal}%
  \BibitemOpen
  \bibfield  {author} {\bibinfo {author} {\bibfnamefont {Y.}~\bibnamefont
  {An}}, \bibinfo {author} {\bibfnamefont {A.}~\bibnamefont {Behnam}}, \bibinfo
  {author} {\bibfnamefont {E.}~\bibnamefont {Pop}}, \ and\ \bibinfo {author}
  {\bibfnamefont {A.}~\bibnamefont {Ural}},\ }\href@noop {} {\bibfield
  {journal} {\bibinfo  {journal} {Applied Physics Letters}\ }\textbf {\bibinfo
  {volume} {102}},\ \bibinfo {pages} {013110} (\bibinfo {year}
  {2013}{\natexlab{b}})}\BibitemShut {NoStop}%
\bibitem [{\citenamefont {An}\ \emph {et~al.}(2015)\citenamefont {An},
  \citenamefont {Behnam}, \citenamefont {Pop}, \citenamefont {Bosman},\ and\
  \citenamefont {Ural}}]{an2015forward}%
  \BibitemOpen
  \bibfield  {author} {\bibinfo {author} {\bibfnamefont {Y.}~\bibnamefont
  {An}}, \bibinfo {author} {\bibfnamefont {A.}~\bibnamefont {Behnam}}, \bibinfo
  {author} {\bibfnamefont {E.}~\bibnamefont {Pop}}, \bibinfo {author}
  {\bibfnamefont {G.}~\bibnamefont {Bosman}}, \ and\ \bibinfo {author}
  {\bibfnamefont {A.}~\bibnamefont {Ural}},\ }\href@noop {} {\bibfield
  {journal} {\bibinfo  {journal} {Journal of Applied Physics}\ }\textbf
  {\bibinfo {volume} {118}},\ \bibinfo {pages} {114307} (\bibinfo {year}
  {2015})}\BibitemShut {NoStop}%
\bibitem [{\citenamefont {Jiao}\ \emph {et~al.}(2016)\citenamefont {Jiao},
  \citenamefont {Wei}, \citenamefont {Song}, \citenamefont {Sun}, \citenamefont
  {Yang}, \citenamefont {Yu}, \citenamefont {Feng}, \citenamefont {Sun},
  \citenamefont {Wei}, \citenamefont {Shi} \emph {et~al.}}]{jiao2016high}%
  \BibitemOpen
  \bibfield  {author} {\bibinfo {author} {\bibfnamefont {T.}~\bibnamefont
  {Jiao}}, \bibinfo {author} {\bibfnamefont {D.}~\bibnamefont {Wei}}, \bibinfo
  {author} {\bibfnamefont {X.}~\bibnamefont {Song}}, \bibinfo {author}
  {\bibfnamefont {T.}~\bibnamefont {Sun}}, \bibinfo {author} {\bibfnamefont
  {J.}~\bibnamefont {Yang}}, \bibinfo {author} {\bibfnamefont {L.}~\bibnamefont
  {Yu}}, \bibinfo {author} {\bibfnamefont {Y.}~\bibnamefont {Feng}}, \bibinfo
  {author} {\bibfnamefont {W.}~\bibnamefont {Sun}}, \bibinfo {author}
  {\bibfnamefont {W.}~\bibnamefont {Wei}}, \bibinfo {author} {\bibfnamefont
  {H.}~\bibnamefont {Shi}},  \emph {et~al.},\ }\href@noop {} {\bibfield
  {journal} {\bibinfo  {journal} {RSC Advances}\ }\textbf {\bibinfo {volume}
  {6}},\ \bibinfo {pages} {10175} (\bibinfo {year} {2016})}\BibitemShut
  {NoStop}%
\bibitem [{\citenamefont {Lin}(2015)}]{lin2015correlation}%
  \BibitemOpen
  \bibfield  {author} {\bibinfo {author} {\bibfnamefont {Y.-J.}\ \bibnamefont
  {Lin}},\ }\href@noop {} {\bibfield  {journal} {\bibinfo  {journal}
  {Superlattices and Microstructures}\ }\textbf {\bibinfo {volume} {88}},\
  \bibinfo {pages} {645} (\bibinfo {year} {2015})}\BibitemShut {NoStop}%
\bibitem [{\citenamefont {Yim}, \citenamefont {McEvoy},\ and\ \citenamefont
  {Duesberg}(2013)}]{yim2013characterization}%
  \BibitemOpen
  \bibfield  {author} {\bibinfo {author} {\bibfnamefont {C.}~\bibnamefont
  {Yim}}, \bibinfo {author} {\bibfnamefont {N.}~\bibnamefont {McEvoy}}, \ and\
  \bibinfo {author} {\bibfnamefont {G.~S.}\ \bibnamefont {Duesberg}},\
  }\href@noop {} {\bibfield  {journal} {\bibinfo  {journal} {Applied Physics
  Letters}\ }\textbf {\bibinfo {volume} {103}},\ \bibinfo {pages} {193106}
  (\bibinfo {year} {2013})}\BibitemShut {NoStop}%
\bibitem [{\citenamefont {Parui}\ \emph {et~al.}(2014)\citenamefont {Parui},
  \citenamefont {Ruiter}, \citenamefont {Zomer}, \citenamefont {Wojtaszek},
  \citenamefont {Van~Wees},\ and\ \citenamefont
  {Banerjee}}]{parui2014temperature}%
  \BibitemOpen
  \bibfield  {author} {\bibinfo {author} {\bibfnamefont {S.}~\bibnamefont
  {Parui}}, \bibinfo {author} {\bibfnamefont {R.}~\bibnamefont {Ruiter}},
  \bibinfo {author} {\bibfnamefont {P.}~\bibnamefont {Zomer}}, \bibinfo
  {author} {\bibfnamefont {M.}~\bibnamefont {Wojtaszek}}, \bibinfo {author}
  {\bibfnamefont {B.}~\bibnamefont {Van~Wees}}, \ and\ \bibinfo {author}
  {\bibfnamefont {T.}~\bibnamefont {Banerjee}},\ }\href@noop {} {\bibfield
  {journal} {\bibinfo  {journal} {Journal of Applied Physics}\ }\textbf
  {\bibinfo {volume} {116}},\ \bibinfo {pages} {244505} (\bibinfo {year}
  {2014})}\BibitemShut {NoStop}%
\bibitem [{\citenamefont {Wang}\ \emph {et~al.}(2013)\citenamefont {Wang},
  \citenamefont {Cheng}, \citenamefont {Xu}, \citenamefont {Tsang},\ and\
  \citenamefont {Xu}}]{wang2013high}%
  \BibitemOpen
  \bibfield  {author} {\bibinfo {author} {\bibfnamefont {X.}~\bibnamefont
  {Wang}}, \bibinfo {author} {\bibfnamefont {Z.}~\bibnamefont {Cheng}},
  \bibinfo {author} {\bibfnamefont {K.}~\bibnamefont {Xu}}, \bibinfo {author}
  {\bibfnamefont {H.~K.}\ \bibnamefont {Tsang}}, \ and\ \bibinfo {author}
  {\bibfnamefont {J.-B.}\ \bibnamefont {Xu}},\ }\href@noop {} {\bibfield
  {journal} {\bibinfo  {journal} {Nature Photonics}\ }\textbf {\bibinfo
  {volume} {7}},\ \bibinfo {pages} {888} (\bibinfo {year} {2013})}\BibitemShut
  {NoStop}%
\bibitem [{\citenamefont {Amirmazlaghani}\ \emph {et~al.}(2013)\citenamefont
  {Amirmazlaghani}, \citenamefont {Raissi}, \citenamefont {Habibpour},
  \citenamefont {Vukusic},\ and\ \citenamefont
  {Stake}}]{amirmazlaghani2013graphene}%
  \BibitemOpen
  \bibfield  {author} {\bibinfo {author} {\bibfnamefont {M.}~\bibnamefont
  {Amirmazlaghani}}, \bibinfo {author} {\bibfnamefont {F.}~\bibnamefont
  {Raissi}}, \bibinfo {author} {\bibfnamefont {O.}~\bibnamefont {Habibpour}},
  \bibinfo {author} {\bibfnamefont {J.}~\bibnamefont {Vukusic}}, \ and\
  \bibinfo {author} {\bibfnamefont {J.}~\bibnamefont {Stake}},\ }\href@noop {}
  {\bibfield  {journal} {\bibinfo  {journal} {IEEE Journal of Quantum
  Electronics}\ }\textbf {\bibinfo {volume} {49}},\ \bibinfo {pages} {589}
  (\bibinfo {year} {2013})}\BibitemShut {NoStop}%
\bibitem [{\citenamefont {Lv}\ \emph {et~al.}(2013)\citenamefont {Lv},
  \citenamefont {Zhang}, \citenamefont {Zhang}, \citenamefont {Deng},\ and\
  \citenamefont {Jie}}]{lv2013high}%
  \BibitemOpen
  \bibfield  {author} {\bibinfo {author} {\bibfnamefont {P.}~\bibnamefont
  {Lv}}, \bibinfo {author} {\bibfnamefont {X.}~\bibnamefont {Zhang}}, \bibinfo
  {author} {\bibfnamefont {X.}~\bibnamefont {Zhang}}, \bibinfo {author}
  {\bibfnamefont {W.}~\bibnamefont {Deng}}, \ and\ \bibinfo {author}
  {\bibfnamefont {J.}~\bibnamefont {Jie}},\ }\href@noop {} {\bibfield
  {journal} {\bibinfo  {journal} {IEEE Electron Device Letters}\ }\textbf
  {\bibinfo {volume} {34}},\ \bibinfo {pages} {1337} (\bibinfo {year}
  {2013})}\BibitemShut {NoStop}%
\bibitem [{\citenamefont {Chen}\ \emph {et~al.}(2015)\citenamefont {Chen},
  \citenamefont {Cheng}, \citenamefont {Wang}, \citenamefont {Wan},
  \citenamefont {Shu}, \citenamefont {Tsang}, \citenamefont {Ho},\ and\
  \citenamefont {Xu}}]{chen2015high}%
  \BibitemOpen
  \bibfield  {author} {\bibinfo {author} {\bibfnamefont {Z.}~\bibnamefont
  {Chen}}, \bibinfo {author} {\bibfnamefont {Z.}~\bibnamefont {Cheng}},
  \bibinfo {author} {\bibfnamefont {J.}~\bibnamefont {Wang}}, \bibinfo {author}
  {\bibfnamefont {X.}~\bibnamefont {Wan}}, \bibinfo {author} {\bibfnamefont
  {C.}~\bibnamefont {Shu}}, \bibinfo {author} {\bibfnamefont {H.~K.}\
  \bibnamefont {Tsang}}, \bibinfo {author} {\bibfnamefont {H.~P.}\ \bibnamefont
  {Ho}}, \ and\ \bibinfo {author} {\bibfnamefont {J.-B.}\ \bibnamefont {Xu}},\
  }\href@noop {} {\bibfield  {journal} {\bibinfo  {journal} {Advanced Optical
  Materials}\ }\textbf {\bibinfo {volume} {3}},\ \bibinfo {pages} {1207}
  (\bibinfo {year} {2015})}\BibitemShut {NoStop}%
\bibitem [{\citenamefont {Riazimehr}\ \emph {et~al.}(2016)\citenamefont
  {Riazimehr}, \citenamefont {Bablich}, \citenamefont {Schneider},
  \citenamefont {Kataria}, \citenamefont {Passi}, \citenamefont {Yim},
  \citenamefont {Duesberg},\ and\ \citenamefont
  {Lemme}}]{riazimehr2016spectral}%
  \BibitemOpen
  \bibfield  {author} {\bibinfo {author} {\bibfnamefont {S.}~\bibnamefont
  {Riazimehr}}, \bibinfo {author} {\bibfnamefont {A.}~\bibnamefont {Bablich}},
  \bibinfo {author} {\bibfnamefont {D.}~\bibnamefont {Schneider}}, \bibinfo
  {author} {\bibfnamefont {S.}~\bibnamefont {Kataria}}, \bibinfo {author}
  {\bibfnamefont {V.}~\bibnamefont {Passi}}, \bibinfo {author} {\bibfnamefont
  {C.}~\bibnamefont {Yim}}, \bibinfo {author} {\bibfnamefont {G.~S.}\
  \bibnamefont {Duesberg}}, \ and\ \bibinfo {author} {\bibfnamefont {M.~C.}\
  \bibnamefont {Lemme}},\ }\href@noop {} {\bibfield  {journal} {\bibinfo
  {journal} {Solid-State Electronics}\ }\textbf {\bibinfo {volume} {115}},\
  \bibinfo {pages} {207} (\bibinfo {year} {2016})}\BibitemShut {NoStop}%
\bibitem [{\citenamefont {Riazimehr}\ \emph {et~al.}(2017)\citenamefont
  {Riazimehr}, \citenamefont {Kataria}, \citenamefont {Bornemann},
  \citenamefont {Haring~Bolívar}, \citenamefont {Ruiz}, \citenamefont
  {Engström}, \citenamefont {Godoy},\ and\ \citenamefont
  {Lemme}}]{riazimehr2017high}%
  \BibitemOpen
  \bibfield  {author} {\bibinfo {author} {\bibfnamefont {S.}~\bibnamefont
  {Riazimehr}}, \bibinfo {author} {\bibfnamefont {S.}~\bibnamefont {Kataria}},
  \bibinfo {author} {\bibfnamefont {R.}~\bibnamefont {Bornemann}}, \bibinfo
  {author} {\bibfnamefont {P.}~\bibnamefont {Haring~Bolívar}}, \bibinfo
  {author} {\bibfnamefont {F.~J.~G.}\ \bibnamefont {Ruiz}}, \bibinfo {author}
  {\bibfnamefont {O.}~\bibnamefont {Engström}}, \bibinfo {author}
  {\bibfnamefont {A.}~\bibnamefont {Godoy}}, \ and\ \bibinfo {author}
  {\bibfnamefont {M.~C.}\ \bibnamefont {Lemme}},\ }\href@noop {} {\bibfield
  {journal} {\bibinfo  {journal} {ACS Photonics}\ }\textbf {\bibinfo {volume}
  {4}},\ \bibinfo {pages} {1506} (\bibinfo {year} {2017})}\BibitemShut
  {NoStop}%
\bibitem [{\citenamefont {Shen}\ \emph {et~al.}(2017)\citenamefont {Shen},
  \citenamefont {Liu}, \citenamefont {Song}, \citenamefont {Li}, \citenamefont
  {Wang}, \citenamefont {Zhou}, \citenamefont {Luo}, \citenamefont {Feng},
  \citenamefont {Wei}, \citenamefont {Lu} \emph {et~al.}}]{shen2017high}%
  \BibitemOpen
  \bibfield  {author} {\bibinfo {author} {\bibfnamefont {J.}~\bibnamefont
  {Shen}}, \bibinfo {author} {\bibfnamefont {X.}~\bibnamefont {Liu}}, \bibinfo
  {author} {\bibfnamefont {X.}~\bibnamefont {Song}}, \bibinfo {author}
  {\bibfnamefont {X.}~\bibnamefont {Li}}, \bibinfo {author} {\bibfnamefont
  {J.}~\bibnamefont {Wang}}, \bibinfo {author} {\bibfnamefont {Q.}~\bibnamefont
  {Zhou}}, \bibinfo {author} {\bibfnamefont {S.}~\bibnamefont {Luo}}, \bibinfo
  {author} {\bibfnamefont {W.}~\bibnamefont {Feng}}, \bibinfo {author}
  {\bibfnamefont {X.}~\bibnamefont {Wei}}, \bibinfo {author} {\bibfnamefont
  {S.}~\bibnamefont {Lu}},  \emph {et~al.},\ }\href@noop {} {\bibfield
  {journal} {\bibinfo  {journal} {Nanoscale}\ }\textbf {\bibinfo {volume}
  {9}},\ \bibinfo {pages} {6020} (\bibinfo {year} {2017})}\BibitemShut
  {NoStop}%
\bibitem [{\citenamefont {Di~Bartolomeo}\ \emph {et~al.}(2017)\citenamefont
  {Di~Bartolomeo}, \citenamefont {Luongo}, \citenamefont {Giubileo},
  \citenamefont {Funicello}, \citenamefont {Niu}, \citenamefont {Schroeder},
  \citenamefont {Lisker},\ and\ \citenamefont {Lupina}}]{di2017hybrid}%
  \BibitemOpen
  \bibfield  {author} {\bibinfo {author} {\bibfnamefont {A.}~\bibnamefont
  {Di~Bartolomeo}}, \bibinfo {author} {\bibfnamefont {G.}~\bibnamefont
  {Luongo}}, \bibinfo {author} {\bibfnamefont {F.}~\bibnamefont {Giubileo}},
  \bibinfo {author} {\bibfnamefont {N.}~\bibnamefont {Funicello}}, \bibinfo
  {author} {\bibfnamefont {G.}~\bibnamefont {Niu}}, \bibinfo {author}
  {\bibfnamefont {T.}~\bibnamefont {Schroeder}}, \bibinfo {author}
  {\bibfnamefont {M.}~\bibnamefont {Lisker}}, \ and\ \bibinfo {author}
  {\bibfnamefont {G.}~\bibnamefont {Lupina}},\ }\href@noop {} {\bibfield
  {journal} {\bibinfo  {journal} {2D Materials}\ }\textbf {\bibinfo {volume}
  {4}},\ \bibinfo {pages} {025075} (\bibinfo {year} {2017})}\BibitemShut
  {NoStop}%
\bibitem [{\citenamefont {Selvi}\ \emph {et~al.}(2018)\citenamefont {Selvi},
  \citenamefont {Unsuree}, \citenamefont {Whittaker}, \citenamefont {Halsall},
  \citenamefont {Hill}, \citenamefont {Thomas}, \citenamefont {Parkinson},\
  and\ \citenamefont {Echtermeyer}}]{C7NR09591K}%
  \BibitemOpen
  \bibfield  {author} {\bibinfo {author} {\bibfnamefont {H.}~\bibnamefont
  {Selvi}}, \bibinfo {author} {\bibfnamefont {N.}~\bibnamefont {Unsuree}},
  \bibinfo {author} {\bibfnamefont {E.}~\bibnamefont {Whittaker}}, \bibinfo
  {author} {\bibfnamefont {M.~P.}\ \bibnamefont {Halsall}}, \bibinfo {author}
  {\bibfnamefont {E.~W.}\ \bibnamefont {Hill}}, \bibinfo {author}
  {\bibfnamefont {A.}~\bibnamefont {Thomas}}, \bibinfo {author} {\bibfnamefont
  {P.}~\bibnamefont {Parkinson}}, \ and\ \bibinfo {author} {\bibfnamefont
  {T.~J.}\ \bibnamefont {Echtermeyer}},\ }\href@noop {} {\bibfield  {journal}
  {\bibinfo  {journal} {Nanoscale}\ }\textbf {\bibinfo {volume} {10}},\
  \bibinfo {pages} {3399} (\bibinfo {year} {2018})}\BibitemShut {NoStop}%
\bibitem [{\citenamefont {Kim}\ \emph {et~al.}(2016)\citenamefont {Kim},
  \citenamefont {Li}, \citenamefont {Chaves}, \citenamefont {Jim{\'e}nez},
  \citenamefont {Rodriguez}, \citenamefont {Susoma}, \citenamefont {Fenner},
  \citenamefont {Lipsanen},\ and\ \citenamefont {Riikonen}}]{kim2016tunable}%
  \BibitemOpen
  \bibfield  {author} {\bibinfo {author} {\bibfnamefont {W.}~\bibnamefont
  {Kim}}, \bibinfo {author} {\bibfnamefont {C.}~\bibnamefont {Li}}, \bibinfo
  {author} {\bibfnamefont {F.~A.}\ \bibnamefont {Chaves}}, \bibinfo {author}
  {\bibfnamefont {D.}~\bibnamefont {Jim{\'e}nez}}, \bibinfo {author}
  {\bibfnamefont {R.~D.}\ \bibnamefont {Rodriguez}}, \bibinfo {author}
  {\bibfnamefont {J.}~\bibnamefont {Susoma}}, \bibinfo {author} {\bibfnamefont
  {M.~A.}\ \bibnamefont {Fenner}}, \bibinfo {author} {\bibfnamefont
  {H.}~\bibnamefont {Lipsanen}}, \ and\ \bibinfo {author} {\bibfnamefont
  {J.}~\bibnamefont {Riikonen}},\ }\href@noop {} {\bibfield  {journal}
  {\bibinfo  {journal} {Advanced Materials}\ }\textbf {\bibinfo {volume}
  {28}},\ \bibinfo {pages} {1845} (\bibinfo {year} {2016})}\BibitemShut
  {NoStop}%
\bibitem [{\citenamefont {Tomer}\ \emph {et~al.}(2015)\citenamefont {Tomer},
  \citenamefont {Rajput}, \citenamefont {Hudy}, \citenamefont {Li},\ and\
  \citenamefont {Li}}]{tomer2015carrier}%
  \BibitemOpen
  \bibfield  {author} {\bibinfo {author} {\bibfnamefont {D.}~\bibnamefont
  {Tomer}}, \bibinfo {author} {\bibfnamefont {S.}~\bibnamefont {Rajput}},
  \bibinfo {author} {\bibfnamefont {L.}~\bibnamefont {Hudy}}, \bibinfo {author}
  {\bibfnamefont {C.}~\bibnamefont {Li}}, \ and\ \bibinfo {author}
  {\bibfnamefont {L.}~\bibnamefont {Li}},\ }\href@noop {} {\bibfield  {journal}
  {\bibinfo  {journal} {Applied Physics Letters}\ }\textbf {\bibinfo {volume}
  {106}},\ \bibinfo {pages} {173510} (\bibinfo {year} {2015})}\BibitemShut
  {NoStop}%
\bibitem [{\citenamefont {Kumar}\ \emph {et~al.}(2016)\citenamefont {Kumar},
  \citenamefont {Kashid}, \citenamefont {Ghosh}, \citenamefont {Kumar},\ and\
  \citenamefont {Singh}}]{kumar2016enhanced}%
  \BibitemOpen
  \bibfield  {author} {\bibinfo {author} {\bibfnamefont {A.}~\bibnamefont
  {Kumar}}, \bibinfo {author} {\bibfnamefont {R.}~\bibnamefont {Kashid}},
  \bibinfo {author} {\bibfnamefont {A.}~\bibnamefont {Ghosh}}, \bibinfo
  {author} {\bibfnamefont {V.}~\bibnamefont {Kumar}}, \ and\ \bibinfo {author}
  {\bibfnamefont {R.}~\bibnamefont {Singh}},\ }\href@noop {} {\bibfield
  {journal} {\bibinfo  {journal} {ACS Applied Materials \& Interfaces}\
  }\textbf {\bibinfo {volume} {8}},\ \bibinfo {pages} {8213} (\bibinfo {year}
  {2016})}\BibitemShut {NoStop}%
\bibitem [{\citenamefont {Mills}\ \emph {et~al.}(2015)\citenamefont {Mills},
  \citenamefont {Min}, \citenamefont {Kim}, \citenamefont {Kim}, \citenamefont
  {Kang}, \citenamefont {Song}, \citenamefont {Myung}, \citenamefont {Lim},
  \citenamefont {An}, \citenamefont {Jung} \emph {et~al.}}]{mills2015direct}%
  \BibitemOpen
  \bibfield  {author} {\bibinfo {author} {\bibfnamefont {E.~M.}\ \bibnamefont
  {Mills}}, \bibinfo {author} {\bibfnamefont {B.~K.}\ \bibnamefont {Min}},
  \bibinfo {author} {\bibfnamefont {S.~K.}\ \bibnamefont {Kim}}, \bibinfo
  {author} {\bibfnamefont {S.~J.}\ \bibnamefont {Kim}}, \bibinfo {author}
  {\bibfnamefont {M.-A.}\ \bibnamefont {Kang}}, \bibinfo {author}
  {\bibfnamefont {W.}~\bibnamefont {Song}}, \bibinfo {author} {\bibfnamefont
  {S.}~\bibnamefont {Myung}}, \bibinfo {author} {\bibfnamefont
  {J.}~\bibnamefont {Lim}}, \bibinfo {author} {\bibfnamefont {K.-S.}\
  \bibnamefont {An}}, \bibinfo {author} {\bibfnamefont {J.}~\bibnamefont
  {Jung}},  \emph {et~al.},\ }\href@noop {} {\bibfield  {journal} {\bibinfo
  {journal} {ACS Applied Materials \& Interfaces}\ }\textbf {\bibinfo {volume}
  {7}},\ \bibinfo {pages} {18300} (\bibinfo {year} {2015})}\BibitemShut
  {NoStop}%
\bibitem [{\citenamefont {Khurelbaatar}\ \emph {et~al.}(2015)\citenamefont
  {Khurelbaatar}, \citenamefont {Kil}, \citenamefont {Shim}, \citenamefont
  {Cho}, \citenamefont {Kim}, \citenamefont {Kim},\ and\ \citenamefont
  {Choi}}]{khurelbaatar2015temperature}%
  \BibitemOpen
  \bibfield  {author} {\bibinfo {author} {\bibfnamefont {Z.}~\bibnamefont
  {Khurelbaatar}}, \bibinfo {author} {\bibfnamefont {Y.-H.}\ \bibnamefont
  {Kil}}, \bibinfo {author} {\bibfnamefont {K.-H.}\ \bibnamefont {Shim}},
  \bibinfo {author} {\bibfnamefont {H.}~\bibnamefont {Cho}}, \bibinfo {author}
  {\bibfnamefont {M.-J.}\ \bibnamefont {Kim}}, \bibinfo {author} {\bibfnamefont
  {Y.-T.}\ \bibnamefont {Kim}}, \ and\ \bibinfo {author} {\bibfnamefont
  {C.-J.}\ \bibnamefont {Choi}},\ }\href@noop {} {\bibfield  {journal}
  {\bibinfo  {journal} {Journal of Semiconductor Technology and Science}\
  }\textbf {\bibinfo {volume} {15}},\ \bibinfo {pages} {7} (\bibinfo {year}
  {2015})}\BibitemShut {NoStop}%
\bibitem [{\citenamefont {Poudel}\ \emph {et~al.}(2017)\citenamefont {Poudel},
  \citenamefont {Liang}, \citenamefont {Choi}, \citenamefont {Hou},
  \citenamefont {Shen}, \citenamefont {Shi}, \citenamefont {Ang}, \citenamefont
  {Shi},\ and\ \citenamefont {Cronin}}]{poudel2017cross}%
  \BibitemOpen
  \bibfield  {author} {\bibinfo {author} {\bibfnamefont {N.}~\bibnamefont
  {Poudel}}, \bibinfo {author} {\bibfnamefont {S.-J.}\ \bibnamefont {Liang}},
  \bibinfo {author} {\bibfnamefont {D.}~\bibnamefont {Choi}}, \bibinfo {author}
  {\bibfnamefont {B.}~\bibnamefont {Hou}}, \bibinfo {author} {\bibfnamefont
  {L.}~\bibnamefont {Shen}}, \bibinfo {author} {\bibfnamefont {H.}~\bibnamefont
  {Shi}}, \bibinfo {author} {\bibfnamefont {L.~K.}\ \bibnamefont {Ang}},
  \bibinfo {author} {\bibfnamefont {L.}~\bibnamefont {Shi}}, \ and\ \bibinfo
  {author} {\bibfnamefont {S.}~\bibnamefont {Cronin}},\ }\href@noop {}
  {\bibfield  {journal} {\bibinfo  {journal} {Scientific Reports}\ }\textbf
  {\bibinfo {volume} {7}},\ \bibinfo {pages} {14148} (\bibinfo {year}
  {2017})}\BibitemShut {NoStop}%
\bibitem [{\citenamefont {Di~Bartolomeo}(2016)}]{di2016graphene}%
  \BibitemOpen
  \bibfield  {author} {\bibinfo {author} {\bibfnamefont {A.}~\bibnamefont
  {Di~Bartolomeo}},\ }\href@noop {} {\bibfield  {journal} {\bibinfo  {journal}
  {Physics Reports}\ }\textbf {\bibinfo {volume} {606}},\ \bibinfo {pages} {1}
  (\bibinfo {year} {2016})}\BibitemShut {NoStop}%
\bibitem [{\citenamefont {Xu}\ \emph {et~al.}(2016{\natexlab{a}})\citenamefont
  {Xu}, \citenamefont {Yu}, \citenamefont {Yang},\ and\ \citenamefont
  {Yang}}]{xu2016interface}%
  \BibitemOpen
  \bibfield  {author} {\bibinfo {author} {\bibfnamefont {D.}~\bibnamefont
  {Xu}}, \bibinfo {author} {\bibfnamefont {X.}~\bibnamefont {Yu}}, \bibinfo
  {author} {\bibfnamefont {L.}~\bibnamefont {Yang}}, \ and\ \bibinfo {author}
  {\bibfnamefont {D.}~\bibnamefont {Yang}},\ }\href@noop {} {\bibfield
  {journal} {\bibinfo  {journal} {Superlattices and Microstructures}\ }\textbf
  {\bibinfo {volume} {99}},\ \bibinfo {pages} {3} (\bibinfo {year}
  {2016}{\natexlab{a}})}\BibitemShut {NoStop}%
\bibitem [{\citenamefont {Xu}\ \emph {et~al.}(2016{\natexlab{b}})\citenamefont
  {Xu}, \citenamefont {Cheng}, \citenamefont {Du}, \citenamefont {Yang},
  \citenamefont {Yu}, \citenamefont {Luo}, \citenamefont {Yin}, \citenamefont
  {Li}, \citenamefont {Dong}, \citenamefont {Ye} \emph
  {et~al.}}]{xu2016contacts}%
  \BibitemOpen
  \bibfield  {author} {\bibinfo {author} {\bibfnamefont {Y.}~\bibnamefont
  {Xu}}, \bibinfo {author} {\bibfnamefont {C.}~\bibnamefont {Cheng}}, \bibinfo
  {author} {\bibfnamefont {S.}~\bibnamefont {Du}}, \bibinfo {author}
  {\bibfnamefont {J.}~\bibnamefont {Yang}}, \bibinfo {author} {\bibfnamefont
  {B.}~\bibnamefont {Yu}}, \bibinfo {author} {\bibfnamefont {J.}~\bibnamefont
  {Luo}}, \bibinfo {author} {\bibfnamefont {W.}~\bibnamefont {Yin}}, \bibinfo
  {author} {\bibfnamefont {E.}~\bibnamefont {Li}}, \bibinfo {author}
  {\bibfnamefont {S.}~\bibnamefont {Dong}}, \bibinfo {author} {\bibfnamefont
  {P.}~\bibnamefont {Ye}},  \emph {et~al.},\ }\href@noop {} {\bibfield
  {journal} {\bibinfo  {journal} {ACS Nano}\ }\textbf {\bibinfo {volume}
  {10}},\ \bibinfo {pages} {4895} (\bibinfo {year}
  {2016}{\natexlab{b}})}\BibitemShut {NoStop}%
\bibitem [{\citenamefont {Di~Bartolomeo}\ \emph {et~al.}(2016)\citenamefont
  {Di~Bartolomeo}, \citenamefont {Giubileo}, \citenamefont {Luongo},
  \citenamefont {Iemmo}, \citenamefont {Martucciello}, \citenamefont {Niu},
  \citenamefont {Fraschke}, \citenamefont {Skibitzki}, \citenamefont
  {Schroeder},\ and\ \citenamefont {Lupina}}]{di2016tunable}%
  \BibitemOpen
  \bibfield  {author} {\bibinfo {author} {\bibfnamefont {A.}~\bibnamefont
  {Di~Bartolomeo}}, \bibinfo {author} {\bibfnamefont {F.}~\bibnamefont
  {Giubileo}}, \bibinfo {author} {\bibfnamefont {G.}~\bibnamefont {Luongo}},
  \bibinfo {author} {\bibfnamefont {L.}~\bibnamefont {Iemmo}}, \bibinfo
  {author} {\bibfnamefont {N.}~\bibnamefont {Martucciello}}, \bibinfo {author}
  {\bibfnamefont {G.}~\bibnamefont {Niu}}, \bibinfo {author} {\bibfnamefont
  {M.}~\bibnamefont {Fraschke}}, \bibinfo {author} {\bibfnamefont
  {O.}~\bibnamefont {Skibitzki}}, \bibinfo {author} {\bibfnamefont
  {T.}~\bibnamefont {Schroeder}}, \ and\ \bibinfo {author} {\bibfnamefont
  {G.}~\bibnamefont {Lupina}},\ }\href@noop {} {\bibfield  {journal} {\bibinfo
  {journal} {2D Materials}\ }\textbf {\bibinfo {volume} {4}},\ \bibinfo {pages}
  {015024} (\bibinfo {year} {2016})}\BibitemShut {NoStop}%
\bibitem [{\citenamefont {Yawei}\ \emph {et~al.}(2015)\citenamefont {Yawei},
  \citenamefont {Liu}, \citenamefont {Ma}, \citenamefont {Xu},\ and\
  \citenamefont {Zhang}}]{yawei2015two}%
  \BibitemOpen
  \bibfield  {author} {\bibinfo {author} {\bibfnamefont {K.}~\bibnamefont
  {Yawei}}, \bibinfo {author} {\bibfnamefont {Y.}~\bibnamefont {Liu}}, \bibinfo
  {author} {\bibfnamefont {Y.}~\bibnamefont {Ma}}, \bibinfo {author}
  {\bibfnamefont {J.}~\bibnamefont {Xu}}, \ and\ \bibinfo {author}
  {\bibfnamefont {D.}~\bibnamefont {Zhang}},\ }in\ \href@noop {} {\emph
  {\bibinfo {booktitle} {Photonics for Energy}}}\ (\bibinfo {organization}
  {Optical Society of America},\ \bibinfo {year} {2015})\ pp.\ \bibinfo {pages}
  {JW3A--4}\BibitemShut {NoStop}%
\bibitem [{\citenamefont {Olawole}\ and\ \citenamefont
  {De}(2018)}]{olawole2018theoretical}%
  \BibitemOpen
  \bibfield  {author} {\bibinfo {author} {\bibfnamefont {O.~C.}\ \bibnamefont
  {Olawole}}\ and\ \bibinfo {author} {\bibfnamefont {D.~K.}\ \bibnamefont
  {De}},\ }\href@noop {} {\bibfield  {journal} {\bibinfo  {journal} {Journal of
  Photonics for Energy}\ }\textbf {\bibinfo {volume} {8}},\ \bibinfo {pages}
  {018001} (\bibinfo {year} {2018})}\BibitemShut {NoStop}%
\bibitem [{\citenamefont {Miao}\ \emph {et~al.}(2012)\citenamefont {Miao},
  \citenamefont {Tongay}, \citenamefont {Petterson}, \citenamefont {Berke},
  \citenamefont {Rinzler}, \citenamefont {Appleton},\ and\ \citenamefont
  {Hebard}}]{miao2012high}%
  \BibitemOpen
  \bibfield  {author} {\bibinfo {author} {\bibfnamefont {X.}~\bibnamefont
  {Miao}}, \bibinfo {author} {\bibfnamefont {S.}~\bibnamefont {Tongay}},
  \bibinfo {author} {\bibfnamefont {M.~K.}\ \bibnamefont {Petterson}}, \bibinfo
  {author} {\bibfnamefont {K.}~\bibnamefont {Berke}}, \bibinfo {author}
  {\bibfnamefont {A.~G.}\ \bibnamefont {Rinzler}}, \bibinfo {author}
  {\bibfnamefont {B.~R.}\ \bibnamefont {Appleton}}, \ and\ \bibinfo {author}
  {\bibfnamefont {A.~F.}\ \bibnamefont {Hebard}},\ }\href@noop {} {\bibfield
  {journal} {\bibinfo  {journal} {Nano Letters}\ }\textbf {\bibinfo {volume}
  {12}},\ \bibinfo {pages} {2745} (\bibinfo {year} {2012})}\BibitemShut
  {NoStop}%
\bibitem [{\citenamefont {An}, \citenamefont {Liu},\ and\ \citenamefont
  {Kar}(2013)}]{an2013optimizing}%
  \BibitemOpen
  \bibfield  {author} {\bibinfo {author} {\bibfnamefont {X.}~\bibnamefont
  {An}}, \bibinfo {author} {\bibfnamefont {F.}~\bibnamefont {Liu}}, \ and\
  \bibinfo {author} {\bibfnamefont {S.}~\bibnamefont {Kar}},\ }\href@noop {}
  {\bibfield  {journal} {\bibinfo  {journal} {Carbon}\ }\textbf {\bibinfo
  {volume} {57}},\ \bibinfo {pages} {329} (\bibinfo {year} {2013})}\BibitemShut
  {NoStop}%
\bibitem [{\citenamefont {Li}, \citenamefont {Lv},\ and\ \citenamefont
  {Zhu}(2015)}]{li2015carbon}%
  \BibitemOpen
  \bibfield  {author} {\bibinfo {author} {\bibfnamefont {X.}~\bibnamefont
  {Li}}, \bibinfo {author} {\bibfnamefont {Z.}~\bibnamefont {Lv}}, \ and\
  \bibinfo {author} {\bibfnamefont {H.}~\bibnamefont {Zhu}},\ }\href@noop {}
  {\bibfield  {journal} {\bibinfo  {journal} {Advanced Materials}\ }\textbf
  {\bibinfo {volume} {27}},\ \bibinfo {pages} {6549} (\bibinfo {year}
  {2015})}\BibitemShut {NoStop}%
\bibitem [{\citenamefont {Crowell}\ and\ \citenamefont
  {Sze}(1966)}]{crowell1966current}%
  \BibitemOpen
  \bibfield  {author} {\bibinfo {author} {\bibfnamefont {C.}~\bibnamefont
  {Crowell}}\ and\ \bibinfo {author} {\bibfnamefont {S.}~\bibnamefont {Sze}},\
  }\href@noop {} {\bibfield  {journal} {\bibinfo  {journal} {Solid-State
  Electronics}\ }\textbf {\bibinfo {volume} {9}},\ \bibinfo {pages} {1035}
  (\bibinfo {year} {1966})}\BibitemShut {NoStop}%
\bibitem [{\citenamefont {Varonides}(2016)}]{varonides2016combined}%
  \BibitemOpen
  \bibfield  {author} {\bibinfo {author} {\bibfnamefont {A.}~\bibnamefont
  {Varonides}},\ }\href@noop {} {\bibfield  {journal} {\bibinfo  {journal}
  {physica status solidi (c)}\ }\textbf {\bibinfo {volume} {13}},\ \bibinfo
  {pages} {1040} (\bibinfo {year} {2016})}\BibitemShut {NoStop}%
\bibitem [{\citenamefont {Liang}\ and\ \citenamefont
  {Ang}(2015)}]{liang2015electron}%
  \BibitemOpen
  \bibfield  {author} {\bibinfo {author} {\bibfnamefont {S.-J.}\ \bibnamefont
  {Liang}}\ and\ \bibinfo {author} {\bibfnamefont {L.~K.}\ \bibnamefont
  {Ang}},\ }\href@noop {} {\bibfield  {journal} {\bibinfo  {journal} {Physical
  Review Applied}\ }\textbf {\bibinfo {volume} {3}},\ \bibinfo {pages} {014002}
  (\bibinfo {year} {2015})}\BibitemShut {NoStop}%
\bibitem [{\citenamefont {Ang}\ and\ \citenamefont
  {Ang}(2016)}]{ang2016current}%
  \BibitemOpen
  \bibfield  {author} {\bibinfo {author} {\bibfnamefont {Y.}~\bibnamefont
  {Ang}}\ and\ \bibinfo {author} {\bibfnamefont {L.}~\bibnamefont {Ang}},\
  }\href@noop {} {\bibfield  {journal} {\bibinfo  {journal} {Physical Review
  Applied}\ }\textbf {\bibinfo {volume} {6}},\ \bibinfo {pages} {034013}
  (\bibinfo {year} {2016})}\BibitemShut {NoStop}%
\bibitem [{\citenamefont {Ang}\ \emph {et~al.}(2017)\citenamefont {Ang},
  \citenamefont {Zubair}, \citenamefont {Ooi},\ and\ \citenamefont
  {Ang}}]{ang2017generalized}%
  \BibitemOpen
  \bibfield  {author} {\bibinfo {author} {\bibfnamefont {Y.~S.}\ \bibnamefont
  {Ang}}, \bibinfo {author} {\bibfnamefont {M.}~\bibnamefont {Zubair}},
  \bibinfo {author} {\bibfnamefont {K.}~\bibnamefont {Ooi}}, \ and\ \bibinfo
  {author} {\bibfnamefont {L.}~\bibnamefont {Ang}},\ }\href@noop {} {\bibfield
  {journal} {\bibinfo  {journal} {arXiv preprint arXiv:1711.05898}\ } (\bibinfo
  {year} {2017})}\BibitemShut {NoStop}%
\bibitem [{\citenamefont {Upadhyay~Kahaly}, \citenamefont {Misra},\ and\
  \citenamefont {Mishra}(2017)}]{upadhyay2017photo}%
  \BibitemOpen
  \bibfield  {author} {\bibinfo {author} {\bibfnamefont {M.}~\bibnamefont
  {Upadhyay~Kahaly}}, \bibinfo {author} {\bibfnamefont {S.}~\bibnamefont
  {Misra}}, \ and\ \bibinfo {author} {\bibfnamefont {S.}~\bibnamefont
  {Mishra}},\ }\href@noop {} {\bibfield  {journal} {\bibinfo  {journal}
  {Journal of Applied Physics}\ }\textbf {\bibinfo {volume} {121}},\ \bibinfo
  {pages} {205110} (\bibinfo {year} {2017})}\BibitemShut {NoStop}%
\bibitem [{\citenamefont {Misra}, \citenamefont {Upadhyay~Kahaly},\ and\
  \citenamefont {Mishra}(2017)}]{misra2017thermionic}%
  \BibitemOpen
  \bibfield  {author} {\bibinfo {author} {\bibfnamefont {S.}~\bibnamefont
  {Misra}}, \bibinfo {author} {\bibfnamefont {M.}~\bibnamefont
  {Upadhyay~Kahaly}}, \ and\ \bibinfo {author} {\bibfnamefont {S.}~\bibnamefont
  {Mishra}},\ }\href@noop {} {\bibfield  {journal} {\bibinfo  {journal}
  {Journal of Applied Physics}\ }\textbf {\bibinfo {volume} {121}},\ \bibinfo
  {pages} {065102} (\bibinfo {year} {2017})}\BibitemShut {NoStop}%
\bibitem [{\citenamefont {Wei}, \citenamefont {Chen},\ and\ \citenamefont
  {Peng}(2013)}]{wei2013electron}%
  \BibitemOpen
  \bibfield  {author} {\bibinfo {author} {\bibfnamefont {X.}~\bibnamefont
  {Wei}}, \bibinfo {author} {\bibfnamefont {Q.}~\bibnamefont {Chen}}, \ and\
  \bibinfo {author} {\bibfnamefont {L.}~\bibnamefont {Peng}},\ }\href@noop {}
  {\bibfield  {journal} {\bibinfo  {journal} {AIP Advances}\ }\textbf {\bibinfo
  {volume} {3}},\ \bibinfo {pages} {042130} (\bibinfo {year}
  {2013})}\BibitemShut {NoStop}%
\bibitem [{\citenamefont {Abrikosov}(1988)}]{abrikosov1988fundamentals}%
  \BibitemOpen
  \bibfield  {author} {\bibinfo {author} {\bibfnamefont {A.~A.}\ \bibnamefont
  {Abrikosov}},\ }\href@noop {} {\emph {\bibinfo {title} {Fundamentals of the
  Theory of Metals}}}\ (\bibinfo  {publisher} {New York, NY; Elsevier Science
  Pub. Co. Inc.},\ \bibinfo {year} {1988})\BibitemShut {NoStop}%
\bibitem [{\citenamefont {Lucas}\ and\ \citenamefont
  {Fong}(2018)}]{lucas2018hydrodynamics}%
  \BibitemOpen
  \bibfield  {author} {\bibinfo {author} {\bibfnamefont {A.}~\bibnamefont
  {Lucas}}\ and\ \bibinfo {author} {\bibfnamefont {K.~C.}\ \bibnamefont
  {Fong}},\ }\href@noop {} {\bibfield  {journal} {\bibinfo  {journal} {Journal
  of Physics: Condensed Matter}\ }\textbf {\bibinfo {volume} {30}},\ \bibinfo
  {pages} {053001} (\bibinfo {year} {2018})}\BibitemShut {NoStop}%
\bibitem [{\citenamefont {Landau}\ and\ \citenamefont
  {Lifshitz}(2013)}]{landau2013quantum}%
  \BibitemOpen
  \bibfield  {author} {\bibinfo {author} {\bibfnamefont {L.~D.}\ \bibnamefont
  {Landau}}\ and\ \bibinfo {author} {\bibfnamefont {E.~M.}\ \bibnamefont
  {Lifshitz}},\ }\href@noop {} {\emph {\bibinfo {title} {Quantum Mechanics:
  Non-Relativistic Theory}}},\ Vol.~\bibinfo {volume} {3}\ (\bibinfo
  {publisher} {Elsevier},\ \bibinfo {year} {2013})\BibitemShut {NoStop}%
\bibitem [{\citenamefont {Crowell}(1965)}]{crowell1965richardson}%
  \BibitemOpen
  \bibfield  {author} {\bibinfo {author} {\bibfnamefont {C.}~\bibnamefont
  {Crowell}},\ }\href@noop {} {\bibfield  {journal} {\bibinfo  {journal}
  {Solid-State Electronics}\ }\textbf {\bibinfo {volume} {8}},\ \bibinfo
  {pages} {395} (\bibinfo {year} {1965})}\BibitemShut {NoStop}%
\bibitem [{\citenamefont {Dushman}(1930)}]{dushman1930thermionic}%
  \BibitemOpen
  \bibfield  {author} {\bibinfo {author} {\bibfnamefont {S.}~\bibnamefont
  {Dushman}},\ }\href@noop {} {\bibfield  {journal} {\bibinfo  {journal}
  {Reviews of Modern Physics}\ }\textbf {\bibinfo {volume} {2}},\ \bibinfo
  {pages} {381} (\bibinfo {year} {1930})}\BibitemShut {NoStop}%
\bibitem [{\citenamefont {Missous}\ and\ \citenamefont
  {Rhoderick}(1991)}]{missous1991richardson}%
  \BibitemOpen
  \bibfield  {author} {\bibinfo {author} {\bibfnamefont {M.}~\bibnamefont
  {Missous}}\ and\ \bibinfo {author} {\bibfnamefont {E.}~\bibnamefont
  {Rhoderick}},\ }\href@noop {} {\bibfield  {journal} {\bibinfo  {journal}
  {Journal of Applied Physics}\ }\textbf {\bibinfo {volume} {69}},\ \bibinfo
  {pages} {7142} (\bibinfo {year} {1991})}\BibitemShut {NoStop}%
\bibitem [{\citenamefont {Chen}\ \emph {et~al.}(2012)\citenamefont {Chen},
  \citenamefont {Chang}, \citenamefont {Li}, \citenamefont {Levi},\ and\
  \citenamefont {Cronin}}]{chen2012gate}%
  \BibitemOpen
  \bibfield  {author} {\bibinfo {author} {\bibfnamefont {C.-C.}\ \bibnamefont
  {Chen}}, \bibinfo {author} {\bibfnamefont {C.-C.}\ \bibnamefont {Chang}},
  \bibinfo {author} {\bibfnamefont {Z.}~\bibnamefont {Li}}, \bibinfo {author}
  {\bibfnamefont {A.}~\bibnamefont {Levi}}, \ and\ \bibinfo {author}
  {\bibfnamefont {S.~B.}\ \bibnamefont {Cronin}},\ }\href@noop {} {\bibfield
  {journal} {\bibinfo  {journal} {Applied Physics Letters}\ }\textbf {\bibinfo
  {volume} {101}},\ \bibinfo {pages} {223113} (\bibinfo {year}
  {2012})}\BibitemShut {NoStop}%
\bibitem [{\citenamefont {Malic}, \citenamefont {Winzer},\ and\ \citenamefont
  {Knorr}(2012)}]{malic2012efficient}%
  \BibitemOpen
  \bibfield  {author} {\bibinfo {author} {\bibfnamefont {E.}~\bibnamefont
  {Malic}}, \bibinfo {author} {\bibfnamefont {T.}~\bibnamefont {Winzer}}, \
  and\ \bibinfo {author} {\bibfnamefont {A.}~\bibnamefont {Knorr}},\
  }\href@noop {} {\bibfield  {journal} {\bibinfo  {journal} {Applied Physics
  Letters}\ }\textbf {\bibinfo {volume} {101}},\ \bibinfo {pages} {213110}
  (\bibinfo {year} {2012})}\BibitemShut {NoStop}%
\bibitem [{\citenamefont {Brida}\ \emph {et~al.}(2013)\citenamefont {Brida},
  \citenamefont {Tomadin}, \citenamefont {Manzoni}, \citenamefont {Kim},
  \citenamefont {Lombardo}, \citenamefont {Milana}, \citenamefont {Nair},
  \citenamefont {Novoselov}, \citenamefont {Ferrari}, \citenamefont {Cerullo}
  \emph {et~al.}}]{brida2013ultrafast}%
  \BibitemOpen
  \bibfield  {author} {\bibinfo {author} {\bibfnamefont {D.}~\bibnamefont
  {Brida}}, \bibinfo {author} {\bibfnamefont {A.}~\bibnamefont {Tomadin}},
  \bibinfo {author} {\bibfnamefont {C.}~\bibnamefont {Manzoni}}, \bibinfo
  {author} {\bibfnamefont {Y.~J.}\ \bibnamefont {Kim}}, \bibinfo {author}
  {\bibfnamefont {A.}~\bibnamefont {Lombardo}}, \bibinfo {author}
  {\bibfnamefont {S.}~\bibnamefont {Milana}}, \bibinfo {author} {\bibfnamefont
  {R.~R.}\ \bibnamefont {Nair}}, \bibinfo {author} {\bibfnamefont
  {K.}~\bibnamefont {Novoselov}}, \bibinfo {author} {\bibfnamefont {A.~C.}\
  \bibnamefont {Ferrari}}, \bibinfo {author} {\bibfnamefont {G.}~\bibnamefont
  {Cerullo}},  \emph {et~al.},\ }\href@noop {} {\bibfield  {journal} {\bibinfo
  {journal} {Nature Communications}\ }\textbf {\bibinfo {volume} {4}},\
  \bibinfo {pages} {1987} (\bibinfo {year} {2013})}\BibitemShut {NoStop}%
\bibitem [{\citenamefont {Wu}\ and\ \citenamefont
  {Yang}(1979)}]{wu1979carrier}%
  \BibitemOpen
  \bibfield  {author} {\bibinfo {author} {\bibfnamefont {C.}~\bibnamefont
  {Wu}}\ and\ \bibinfo {author} {\bibfnamefont {E.}~\bibnamefont {Yang}},\
  }\href@noop {} {\bibfield  {journal} {\bibinfo  {journal} {Solid-State
  Electronics}\ }\textbf {\bibinfo {volume} {22}},\ \bibinfo {pages} {241}
  (\bibinfo {year} {1979})}\BibitemShut {NoStop}%
\bibitem [{\citenamefont {Kim}, \citenamefont {Jeong},\ and\ \citenamefont
  {Lundstrom}(2010)}]{kim2010momentum}%
  \BibitemOpen
  \bibfield  {author} {\bibinfo {author} {\bibfnamefont {R.}~\bibnamefont
  {Kim}}, \bibinfo {author} {\bibfnamefont {C.}~\bibnamefont {Jeong}}, \ and\
  \bibinfo {author} {\bibfnamefont {M.~S.}\ \bibnamefont {Lundstrom}},\
  }\href@noop {} {\bibfield  {journal} {\bibinfo  {journal} {Journal of Applied
  Physics}\ }\textbf {\bibinfo {volume} {107}},\ \bibinfo {pages} {054502}
  (\bibinfo {year} {2010})}\BibitemShut {NoStop}%
\end{thebibliography}%
\end{document}